\documentclass[preprintnumbers,amsmath,amssymb,floatfix,12pt,prd,superscriptaddress,nofootinbib]{revtex4}
\usepackage{graphicx}
\usepackage{epsfig}
\usepackage{bm}
\usepackage{amsfonts}
\usepackage{subfigure}

\def\bea{\begin{eqnarray}}
\def\eea{\end{eqnarray}}
\begin{document}
\begin{center}
\LARGE {\bf Cosmological perturbations in warm-tachyon inflationary universe model with viscous pressure}
\end{center}
\begin{center}
{M. R. Setare $^{a}$\footnote{E-mail: rezakord@ipm.ir
}\hspace{1mm} ,
V. Kamali $^{b}$\footnote{E-mail: vkamali1362@gmail.com}\hspace{1.5mm} \\
 $^a$ {\small {\em  Department of Science,  Campus of Bijar, University of Kurdistan, Bijar, Iran}}\hspace{1.5mm}\\
$^b$ {\small {\em  Department of Physics, Faculty of Science,\\
Bu-Ali Sina University, Hamedan, 65178, Iran}}}\\

\end{center}

%\vskip 3cm

\begin{center}
{\bf{Abstract}}\\
We study the warm-tachyon inflationary universe model with viscous pressure in high-dissipation regime.
General conditions which are required for this model to be realizable are derived in the slow-roll approximation.
We present analytic expressions for density perturbation and amplitude of tensor perturbation in longitudinal gauge. Expressions of tensor-to-scalar ratio, scalar spectral index and its running are obtained. We develop our model by using exponential
potential, the characteristics of this model are calculated for two specific cases in
great details: 1- Dissipative parameter $\Gamma$ and bulk viscous parameter $\zeta$ are constant parameters.
2- Dissipative parameter is a function of tachyon field $\phi$ and bulk viscous parameter
is a function of matter-radiation mixture  energy density $\rho$. The parameters of the model are restricted by
 recent observational data from the nine-year Wilkinson microwave
anisotropy probe (WMAP9), Planck and BICEP2 data.
 \end{center}

\newpage

\section{Introduction}
Big Bang model has many long-standing problems (monopole, horizon,
flatness,...). These problems are solved in a framework of
inflationary universe models \cite{1-i}. Scalar field as a source
of inflation provides a causal interpretation of the origin of
the distribution of
Large-Scale Structure (LSS), and also observed anisotropy
of cosmological microwave background (CMB) \cite{6,planck,BICEP2}. The standard
models for inflationary universe are divided
into two regimes, slow-roll and reheating regimes. In the slow-roll
period, kinetic energy remains small compared to the potential
term. In this period, all interactions between scalar fields
(inflatons) and  other fields are neglected and as a result the universe
inflates. Subsequently, in reheating epoch, the kinetic energy
is comparable to the potential energy that causes inflaton to begin an
oscillation around  the minimum of the potential while losing their
energy to other fields present in the theory. After the reheating period, the universe is filled with radiation. \\ In warm inflation scenario the radiation production
occurs during inflationary period and
reheating is avoided \cite{3}. Thermal fluctuations may be
generated during warm inflationary epoch. These fluctuations could play a
dominant role to produce initial fluctuations which are necessary
for Large-Scale Structure (LSS) formation. In this model, density
fluctuation arises from thermal rather than quantum fluctuation
\cite{3-i}. Warm inflationary period ends when the universe stops
inflating. After this period, the universe enters in the radiation
phase smoothly \cite{3}. Finally, remaining inflatons or dominant
radiation fields create matter components of the universe. Some extensions of this model are found in Ref.\cite{new}.\\
In the warm inflation models there has to be continuously particle production.
For this to be possible, then the microscopic processes that produce
these particles must occur at a timescale much faster than Hubble
expansion.  Thus the decay rates $\Gamma_i$ (not to be confused with the
dissipative coefficient) must be bigger than $H$. Also these produced
particles must thermalize. Thus the scattering processes amongst these
produced particles must occur at a rate bigger than $H$.  These
adiabatic conditions were outlined since the early warm inflation
papers, such as Ref. \cite{1-ne}.  More recently
there has been considerable explicit calculation from Quantum Field Theory (QFT) that
explicitly computes all these relevant decay and scattering rates
in warm inflation models \cite{4nn,arj}.\\
In warm inflation models, for simplicity, particles which are created by the inflaton decay are considered as massless particles (or radiation). Existence of massive particles in the inflationary fluid model as a new model of inflation was studied in Ref.\cite{41-i}.
Perturbation parameters of this model were obtained in Ref.\cite{2-ne}.
In this scenario the existence of massive particles  alters the dynamic of the inflationary universe models by modification of the fluid pressure.
Using the random fluid hydrodynamic fluctuation theory which is generalized by Landau and Lifsitz \cite{landa}, we can describe the cosmological fluctuations in the system with radiation and tachyon scalar field.
Decay of the massive particles within the fluid is an entropy-producing scalar phenomenon. In the other hand,
''bulk viscous pressure'' has entropy-producing property.
Therefore, the decay of particles  may be considered by a bulk viscous pressure $\Pi=-3\zeta H$, where $H$ is
Hubble parameter and $\zeta$ is phenomenological coefficient of bulk viscosity \cite{3-ne}.
This coefficient is positive-definite by the second law of thermodynamics and in general depends on the energy density of the fluid.\\
The Friedmann-Robertson-Walker (FRW) cosmological models in the
context of string/M-theory have been related to brane-antibrane
configurations \cite{4-i}. Tachyon fields, associated with
unstable D-branes, are  responsible of inflation in early time
\cite{5-i}. The tachyon inflation is a k-inflation model
\cite{n-1} for scalar field $\phi$ with a positive potential
$V(\phi)$. Tachyon potentials have two special properties,
firstly a maximum of these potential is obtained where
$\phi\rightarrow 0$ and second property is the minimum of these
potentials is obtained where $\phi\rightarrow \infty$. If the
tachyon field starts to roll down the potential, then universe, which is
dominated by a new form of matter, will smoothly evolve from
inflationary universe to an era which is dominated by a
non-relativistic fluid \cite{1}. So, we could explain the phase
of acceleration expansion (inflation) in terms of tachyon field.\\
Cosmological perturbations of warm inflation model (with viscous pressure) have been studied in Ref.\cite{9-f} (\cite{2-ne}).
Warm tachyon inflationary universe model has been studied in
Ref.\cite{1-m}, also warm inflation on the brane (with viscous pressure) has been studied in Ref \cite{6-f} (\cite{v-2}). To the best of our knowledge, a model in which warm tachyon inflation with viscous pressure
has not been yet considered.
In the present work we will study warm-tachyon
inspired inflation with viscous pressure. The paper organized as follow: In the next section, we will
describe warm-tachyon inflationary universe model with viscous pressure and the perturbation parameters
for our model. In section (3), we study our model  using the
exponential potential in high dissipative regime. Finally in
section (5), we close by some concluding  remarks.

\section{The model}
In this section, we will obtain the parameters of the warm tachyon inflation with viscous pressure. This model may be described  by an effective tachyon fluid and matter-radiation imperfect fluid. Tachyon fluid in a spatially flat Friedmann Robertson Walker
(FRW) is recognized by these parameters
\cite{1,v-2}

\begin{eqnarray}\label{2}
T_{\mu}^{\nu}=diag(-\rho_{\phi},P_{\phi},P_{\phi},P_{\phi})\\
\nonumber
P_{\phi}=-V(\phi)\sqrt{1-\dot{\phi}^2},\\
\nonumber
\rho_{\phi}=\frac{V(\phi)}{\sqrt{1-\dot{\phi}^2}},~~~~~~~
\end{eqnarray}
Important characteristics
of the potential are $\frac{dV}{d\phi}<0,$ and $V(\phi\rightarrow
\infty)\rightarrow 0$ \cite{2}. The imperfect fluid is a mixture of matter and radiation of adiabatic index $\gamma$ which has energy density $\rho=Ts(\phi,T)$ ($T$ is temperature  and $s$ is entropy density of the imperfect fluid.) and pressure $P+\Pi$ where, $P=(\gamma-1)\rho$.  $\Pi=-3\zeta H$ is bulk viscous pressure  \cite{3-ne}, where $\zeta$ is phenomenological coefficient of bulk viscosity.
The dynamic of the model in background level  is given by the Friedmann equation,
\begin{eqnarray}\label{4}
3H^2=\rho_T=\frac{V(\phi)}{\sqrt{1-\dot{\phi}^2}}+\rho,
\end{eqnarray}
the conservation equations of tachyon field  and imperfect fluid
\begin{equation}\label{5}
\dot{\rho}_{\phi}+3H(P_{\phi}+\rho_{\phi})=-\Gamma\dot{\phi}^2\Rightarrow
\frac{\ddot{\phi}}{1-\dot{\phi}^2}+3H\dot{\phi}+\frac{V'}{V}=-\frac{\Gamma}{V}\sqrt{1-\dot{\phi}^2}\dot{\phi},
\end{equation}
and
\begin{equation}\label{6}
\dot{\rho}+3H(\rho+P+\Pi)=\dot{\rho}+3H(\gamma\rho+\Pi)=\Gamma\dot{\phi}^2,
\end{equation}
where we have used the natural units ($c=\hbar=1$) and $\frac{8\pi}{m_p^2}=1$.
$\Gamma$ is the dissipative coefficient with the dimension
$m_p^5$. Dissipation term denotes the inflaton decay into the imperfect fluid in the inflationary epoch. In the above equations dots "." mean derivative with
respect to cosmic time, prime  denotes derivative with respect
to the tachyon field $\phi$.
The energy density of radiation  and the entropy density  increase by the bulk viscosity pressure $\Pi$ (see FIG.1 and FIG.2)\cite{v-2}.

During slow-roll inflation epoch the energy density
(\ref{2}) is the order of potential, i.e. $\rho_{\phi}\sim V,$ and
dominates over the imperfect fluid energy density, i.e. $\rho_{\phi}>\rho$.
Using slow-roll approximation when, $\dot{\phi}\ll 1,$ and
$\ddot{\phi}\ll(3H+\frac{\Gamma}{V}),$ \cite{3}  the
dynamic equations (\ref{4}) and (\ref{5})   are reduced to
\begin{eqnarray}\label{7}
3H^2=V~~~~~~~~~,
\\
\nonumber
3H(1+r)\dot{\phi}=-\frac{V'}{V},
\end{eqnarray}
where $r=\frac{\Gamma}{3HV}$. From above equations and
Eq.(\ref{6}),  when the decay of the tachyon field to imperfect fluid is quasi-stable, i.e.  $\dot{\rho}\ll 3H(\gamma\rho+\Pi),$ and $\dot{\rho}\ll\Gamma\dot{\phi}^2$, $\rho$ may be written as
\begin{equation}\label{9}
\rho=\frac{1}{\gamma}(rV\dot{\phi}^2-\Pi)=\frac{1}{\gamma}(\frac{r}{3(1+r)^2}(\frac{V'}{V})^2-\Pi).
\end{equation}

In the present work, we will restrict our analysis in high dissipative regime, i.e. $r\gg 1,$ where the dissipation coefficient $\Gamma$ is much greater than $3HV$ \cite{v-2}. Dissipation parameter $\Gamma$ may be a constant parameter or a positive function of inflaton $\phi$ by the second law of thermodynamics. There are some specific forms for the dissipative coefficient, with the most common which are found in the literatures being the $\Gamma\sim T^3$ form \cite{mm-1},\cite{2nn},\cite{3nn},\cite{4nn}. In some works $\Gamma$ and potential of the inflaton have the same forms \cite{1-m,v-2}. In Ref.\cite{2-ne}, perturbation parameters for warm inflationary model with viscous pressure have been obtained where $\Gamma=\Gamma(\phi)=V(\phi)$ and $\Gamma=\Gamma_0=const$. In this work we will study the warm-tachyon inflationary universe  model with viscous pressure in this two cases. \\
The slow-roll parameters of the model are presented by
\begin{eqnarray}\label{10}
\epsilon=-\frac{\dot{H}}{H^2}\simeq\frac{1}{2(1+r)V}(\frac{V'}{V})^2,
\\
\nonumber
\eta=-\frac{\ddot{H}}{H\dot{H}}\simeq\frac{1}{(1+r)V}(\frac{V''}{V}-\frac{1}{2}(\frac{V'}{V})^2).
\end{eqnarray}
From Eqs.(\ref{9}) and (\ref{10}) we find
\begin{equation}\label{12}
\rho=\frac{1}{\gamma}(\frac{2}{3}\frac{r}{1+r}\epsilon\rho_{\phi}-\Pi).
\end{equation}
The condition of slow-roll is $\epsilon<1$, therefore from above equation,
warm-tachyon inflation with viscous pressure
could take place when
\begin{equation}\label{13}
\rho_{\phi}>\frac{3(1+r)}{2r}[\gamma\rho+\Pi].
\end{equation}
Inflation period ends when, $\epsilon\simeq 1$ which implies
\begin{equation}\label{14}
\rho_{\phi}\simeq\frac{3(1+r)}{2r}[\gamma\rho+\Pi],
\nonumber
[\frac{V'_f}{V_f}]^2\frac{1}{V_f}\simeq 2(1+r_f),
\end{equation}
where the subscript $f$ denotes the end of inflation. The number
of e-folds is given by
\begin{eqnarray}\label{15}
N=\int_{\phi_{*}}^{\phi_f}Hdt=\int_{\phi_{*}}^{\phi_f}\frac{H}{\dot{\phi}}d\phi=-\int_{\phi_{*}}^{\phi_f}\frac{V^2}{V'}(1+r)d\phi.
\end{eqnarray}
where the subscript $*$ denotes the epoch when the cosmological
scale exits the horizon.

We will study inhomogeneous perturbations of the FRW background by using the linear perturbation equation of warm inflation scenario \cite{landa}.
These scalar perturbations in the longitudinal gauge, may be described by the perturbed FRW metric
\begin{equation}\label{16}
ds^2=(1+2\Phi)dt^2-a^2(t)(1-2\Psi)\delta_{ij}dx^idx^j,
\end{equation}
where $\Phi$ and $\Psi$ are gauge-invariant metric perturbation variables \cite{7-f}. All perturbed quantities have a spatial sector of the form $e^{i\mathbf{kx}}$, where $k$ is the wave number. Following Ref.\cite{landa}, we introduce the stress-energy tensor as
\begin{equation}\label{}
T_{ab}=(\rho+P)n_a n_b+Pg_{ab}+n_aq_b+n_bq_b+\Pi_{ab}
\end{equation}
where the trace-free tensor $\Pi_{ab}$ and $q_a$ are orthogonal to the unit vector $n_a$ ($n^{a}$ is the unit normal to the constant-time surface \cite{landa}). For the  linear perturbation theory $\rho$ and $P$ are replaced by $\rho+\delta\rho$ and $P+\delta P$ respectively. We also define the perturbation parameters
\begin{equation}\label{}
q_i=(\rho+P)\nabla_i\delta V~~~~~~~~~~\delta\Pi_{ij}=\nabla_i\nabla_j\delta\Pi-\frac{1}{3}g_{ij}\nabla^2\delta\Pi
\end{equation}

So, the perturbed Einstein field equation equation of motion  in momentum space have these forms
\begin{equation}\label{}
\nonumber
\Phi=\Psi,
\end{equation}
\begin{equation}\label{17}
\dot{\Phi}+H\Phi=\frac{1}{2}[-\frac{(\gamma\rho+\Pi)av}{k}+\frac{V\dot{\phi}}{\sqrt{1-\dot{\phi}^2}}\delta\phi],
\end{equation}
\begin{eqnarray}\label{18}
\frac{\ddot{\delta\phi}}{1-\dot{\phi}^2}+[3H+\frac{\Gamma}{V}]\dot{\delta\phi}+[\frac{k^2}{a^2}+(\ln V)''+\dot{\phi}(\frac{\Gamma}{V})']\delta\phi\\
\nonumber
-[\frac{1}{1-\dot{\phi}^2}+3]\dot{\phi}\dot{\Phi}-[\dot{\phi}\frac{\Gamma}{V}-2(\ln V)']\Phi=0,
\end{eqnarray}
The fluid equations obtain from the stress-energy tensor \cite{landa}.
\begin{eqnarray}\label{19}
(\dot{\delta\rho})+3\gamma H\delta\rho+ka(\gamma\rho+\Pi)v+3(\gamma\rho+\Pi)\dot{\Phi}\\
\nonumber
-\dot{\phi}^2\Gamma'\delta\phi-\Gamma\dot{\phi}[2(\dot{\delta\phi})+\dot{\phi}\Phi]=0,~~~~~~~~~~~~~~
\end{eqnarray}

\begin{equation}\label{20}
\dot{v}+4Hv+\frac{k}{a}[\Phi+\frac{\delta P}{\rho+P}+\frac{\Gamma\dot{\phi}}{\rho+P}\delta\phi]=0.
\end{equation}
where
\begin{equation}\label{}
\nonumber
\delta P=(\gamma-1)\delta\rho+\delta\Pi,~~~~~~~\delta\Pi=\Pi[\frac{\zeta_{,\rho}}{\zeta}\delta\rho+\Phi+\frac{\dot{\Phi}}{H}].
\end{equation}
The above equations are obtained for Fourier components $e^{i\mathbf{kx}}$, where the subscript $k$ is omitted. $v$ in the above set of equations is given by the decomposition of the velocity field ($\delta u_j=-\frac{iak_J}{k}ve^{i\mathbf{kx}}, j=1,2,3$) \cite{6-f}.
Warm inflation model may be considered as a hybrid-like inflationary model where the inflaton field interacts with imperfect fluid \cite{9-f}, \cite{8-f}. Entropy perturbation may be related to dissipation term \cite{10-f}. In slow-roll approximation the set of perturbed equations are reduced to \cite{v-2}
\begin{equation}\label{21}
\Phi\simeq \frac{1}{2H}[-\frac{4(\gamma\rho+\Pi)av}{k}+V\dot{\phi}\delta\phi],
\end{equation}

\begin{equation}\label{22}
[3H+\frac{\Gamma}{V}]\dot{\delta\phi}+[(\ln V)''+\dot{\phi}(\frac{\Gamma}{V})']\delta\phi
\simeq[\dot{\phi}\frac{\Gamma}{V}-2(\ln V)']\Phi,
\end{equation}

\begin{equation}\label{23}
\delta\rho\simeq\frac{\dot{\phi}^2}{3\gamma H}[\Gamma'\delta\phi+\Gamma\Phi],
\end{equation}
and
\begin{eqnarray}\label{24}
v\simeq-\frac{k}{4aH}(\Phi+\frac{(\gamma-1)\delta\rho+\delta\Pi}{\gamma\rho+\Pi}+\frac{\Gamma\dot{\phi}}{\gamma\rho+\Pi}\delta\phi).
\end{eqnarray}
Using Eqs.(\ref{21}), (\ref{23}) and (\ref{24}), perturbation variable $\Phi$ is determined
\begin{eqnarray}\label{25}
\Phi\simeq\frac{\dot{\phi}V}{2H}\frac{\delta\phi}{G(\phi)}[1+\frac{\Gamma}{4HV}+([\gamma-1]+\Pi\frac{\zeta_{,\rho}}{\zeta})\frac{\dot{\phi}\Gamma'}{12\gamma H^2V}],
\end{eqnarray}
where
\begin{eqnarray}\label{}
\nonumber
G(\phi)=1-\frac{1}{8H^2}[2\gamma\rho+3\Pi+\frac{\gamma\rho+\Pi}{\gamma}(\Pi\frac{\zeta_{,\rho}}{\zeta}-1)].
\end{eqnarray}
In Eq.(\ref{25}), for $\Pi\rightarrow 0$ and $\gamma=\frac{4}{3}$ case, we may obtain the perturbation variable $\Phi$ of warm tachyon inflation model without viscous pressure effect \cite{1-m} (In this case, we find $G(\phi)\rightarrow 1$ because of the inequality  $\frac{\rho}{V}\ll 1$.).
 Using Eq.(\ref{7}), we find
\begin{eqnarray}\label{26}
(3H+\frac{\Gamma}{V})\frac{d}{dt}=(3H+\frac{\Gamma}{V})\dot{\phi}\frac{d}{d\phi}=-\frac{V'}{V}\frac{d}{d\phi}.
\end{eqnarray}
From above equation, Eq.(\ref{22}) and Eq.(\ref{25}), the expression $\frac{(\delta\phi)'}{\delta\phi}$ is obtained
\begin{eqnarray}\label{27}
\frac{(\delta\phi)'}{\delta\phi}=\frac{1}{(\ln V)'}[(\ln V)''+\dot{\phi}(\frac{\Gamma}{V})'+(2(\ln V)'-\dot{\phi}\frac{\Gamma}{V})(\frac{V\dot{\phi}}{2GH})\\
\nonumber
\times (1+\frac{\Gamma}{4HV}+[(\gamma-1)+\Pi\frac{\zeta_{,\rho}}{\zeta}]\frac{\dot{\phi}\Gamma'}{12\gamma H^2 V})].~~~~~~~~~~~~~~
\end{eqnarray}
We will return to the above relation soon. Following Refs.\cite{1-m}, \cite{6-f}, \cite{v-2} and  \cite{10-f}, we introduce auxiliary function $\chi$ as
\begin{equation}\label{28}
\chi=\frac{\delta\phi}{(\ln V)'}\exp[\int\frac{1}{3H+\frac{\Gamma}{V}}(\frac{\Gamma}{V})'d\phi].
\end{equation}
From above definition we have
\begin{eqnarray}\label{29}
\frac{\chi'}{\chi}=\frac{(\delta\phi)'}{\delta\phi}-\frac{(\ln V)''}{(\ln V)'}+\frac{(\frac{\Gamma}{V})'}{3H+\frac{\Gamma}{V}}.
\end{eqnarray}
Using above equation, Eqs.(\ref{27})  (\ref{7}) and (\ref{7})
\begin{equation}\label{31}
\frac{\chi'}{\chi}=-\frac{9}{8G}\frac{2H+\frac{\Gamma}{V}}{(3H+\frac{\Gamma}{V})^2}[\Gamma+4HV-([\gamma-1]+\Pi\frac{\zeta_{,\rho}}{\zeta})\frac{\Gamma'(\ln V)'}{3\gamma H(3H+\frac{\Gamma}{V})}]\frac{(\ln V)'}{V}.
\end{equation}
A solution for the above equation is
\begin{eqnarray}\label{32}
\chi(\phi)=C\exp(-\int\{-\frac{9}{8G}\frac{2H+\frac{\Gamma}{V}}{(3H+\frac{\Gamma}{V})^2}~~~~~~~~~~~~~~~~~~~~~~~~~~~~~~~~~~\\
\nonumber \times[\Gamma+4HV-([\gamma-1]+\Pi\frac{\zeta_{,\rho}}{\zeta})\frac{\Gamma'(\ln V)'}{3\gamma H(3H+\frac{\Gamma}{V})}]\frac{(\ln V)'}{V}\}d\phi),
\end{eqnarray}
where $C$ is integration constant. From above equation and Eq.(\ref{29}) we find small change of variable $\delta\phi$
\begin{equation}\label{33}
\delta\phi=C(\ln V)'\exp(\Im(\phi)),
\end{equation}
where
\begin{eqnarray}\label{34}
\Im(\phi)=-\int[\frac{(\frac{\Gamma}{V})'}{3H+\frac{\Gamma}{V}}+\frac{9}{8G}\frac{2H+\frac{\Gamma}{V}}{(3H+\frac{\Gamma}{V})^2}~~~~~~~~~~~~~~~~~~~~~~\\
\nonumber
\times[\Gamma+4HV-([\gamma-1]+\Pi\frac{\zeta_{,\rho}}{\zeta})\frac{\Gamma'(\ln V)'}{3\gamma H(3H+\frac{\Gamma}{V})}]\frac{(\ln V)'}{V}]d\phi
\end{eqnarray}
Finally the density perturbation is given by \cite{12-f}
\begin{equation}\label{35}
\delta_H=\frac{16\pi}{5}\frac{\exp(-\Im(\phi))}{(\ln V)'}\delta\phi=\frac{16\pi}{15}\frac{\exp(-\Im(\phi))}{Hr\dot{\phi}}\delta\phi.
\end{equation}
By inserting $\Gamma=0$ and $\xi=0$, the above equation reduces to $\delta_{H}\simeq\frac{H}{\dot{\phi}}\delta\phi$ which agrees with the density perturbation in cool inflation model \cite{1-i}. In warm inflation model the fluctuations of the scalar field in high dissipative regime ($r\gg 1$) may be generated by thermal fluctuation instead of quantum fluctuations \cite{5} as
\begin{equation}\label{36}
(\delta\phi)^2\simeq\frac{k_F T_r}{2\pi^2},
\end{equation}
where in this limit freeze-out wave number $k_F=\sqrt{\frac{\Gamma H}{V}}=H\sqrt{3r}\geq H$ corresponds to the freeze-out scale at the point when, dissipation damps out to thermally excited fluctuations ($\frac{V''}{V'}<\frac{\Gamma H}{V}$) \cite{5}. With the help of the above equation and Eq.(\ref{35}) in high dissipative regime ($r\gg 1$) we find
\begin{equation}\label{37}
\delta_H^2=\frac{64}{225\sqrt{3}}\frac{\exp(-2\Im(\phi))}{r^{\frac{1}{2}}\tilde{\epsilon}}\frac{T_r}{H},
\end{equation}
where
\begin{equation}\label{38}
\tilde{\Im}(\phi)=-\int[\frac{1}{3Hr}(\frac{\Gamma}{V})'+\frac{9}{8G}(1-[(\gamma-1)+\Pi\frac{\zeta_{,\rho}}{\zeta}]\frac{(\ln\Gamma)'(\ln V)'}{9\gamma rH^2})(\ln V)']d\phi,
\end{equation}
and
\begin{equation}\label{39}
\tilde{\epsilon}=\frac{1}{2 r}\frac{V'^2}{V^3}.
\end{equation}
An important perturbation parameter is scalar index $n_s$ which
in high dissipative regime is given by
\begin{equation}\label{40}
n_s=1+\frac{d\ln \delta_H^2}{d\ln k}\approx
1-\frac{5}{2}\tilde{\epsilon}-\frac{3}{2}\tilde{\eta}+\tilde{\epsilon}(\frac{2V}{V'})(\frac{r'}{4r}-2\tilde{\Im}(\phi)'),
\end{equation}
where
\begin{equation}\label{41}
\tilde{\eta}=\frac{1}{rV}[\frac{V''}{V}-\frac{1}{2}(\frac{V'}{V})^2].
\end{equation}
In Eq.(\ref{40}) we have used a relation between small change of
the number of e-folds and interval in wave number ($dN=-d\ln k$).
The Planck measurement constraints the spectral index as \cite{planck}:
\begin{eqnarray}\label{}
n_s=0.96\pm0.0073
\end{eqnarray}
Running of the scalar spectral index may be found as
\begin{eqnarray}\label{42}
\alpha_s=\frac{dn_s}{d\ln k}=-\frac{dn_s}{dN}=-\frac{d\phi}{dN}\frac{dn_s}{d\phi}=\frac{1}{rV}(\frac{V'}{V})n_s'.
\end{eqnarray}
This parameter is one of the interesting cosmological
perturbation parameters which is approximately $ -0.0134 \pm 0.0090$, by using
Planck observational results \cite{planck}.
\\ During inflation epoch,
there are two independent components of gravitational waves
($h_{\times +}$) with action of massless scalar field are
produced by the generation of tensor perturbations. The amplitude
of tensor perturbation is given by
\begin{eqnarray}\label{43}
A_g^2=2(\frac{H}{2\pi})^2\coth[\frac{k}{2T}]=\frac{V^2}{6\pi^2}\coth[\frac{k}{2T}],
\end{eqnarray}
where, the temperature $T$ in extra factor $\coth[\frac{k}{2T}]$
denotes, the temperature of the thermal background of
gravitational wave \cite{7}. Spectral index $n_g$ may be found as
\begin{eqnarray}\label{44}
n_g=\frac{d}{d\ln k}(\ln [\frac{A_g^2}{\coth(\frac{k}{2T})}])\simeq-2\tilde{\epsilon},
\end{eqnarray}
where $A_g\propto k^{n_g}\coth[\frac{k}{2T}]$ \cite{7}.  Using Eqs.(\ref{37}) and
(\ref{43}) we write the tensor-scalar ratio in high dissipative
regime
\begin{eqnarray}\label{45}
R(k)=\frac{A_g^2}{P_R}|_{k=k_{0}}=\frac{54\sqrt{3}}{5}\frac{r^{\frac{1}{2}}\tilde{\epsilon}H^3}{T_r}\exp(2\Im(\phi))\coth[\frac{k}{2T}]|_{k=k_{0}},
\end{eqnarray}
where $k_{0}$ is referred  to pivot point \cite{7} and $P_R=\frac{25}{4}\delta_H^2$. An upper bound for this parameter is obtained
by using    Planck data, $R<0.11$ \cite{6}.
Non-Gaussianity of the warm-tachyon inflation model is presented in Ref.\cite{1-m} as
\begin{eqnarray}\label{}
f_{NL}=-\frac{5}{3}\frac{\dot{\phi}}{H}[\frac{1}{H}\ln(\frac{k_F}{H})(\frac{V'''+2k_F^2V'}{\Gamma})].
\end{eqnarray}
In high dissipative regime ($r\gg 1$), $f_{NL}$ parameter has the following form
\begin{eqnarray}\label{va}
f_{NL}=\frac{5}{9}(\frac{V'}{V})^2(\frac{\ln r}{r}).
\end{eqnarray}
In the above equation, we have used Eq.(\ref{7}) and definition $k_F=\sqrt{\frac{\Gamma H}{V}}=H\sqrt{3r}$.

We note that, the $\Im(\phi)$ factor (\ref{34})  which is found  in perturbation parameters (\ref{37}), (\ref{40}), (\ref{42}) and (\ref{45}) in high energy limit ($V\gg\lambda$), for tachyonic warm-viscous inflation model  has an important difference with the same factor which was obtained for non-viscous   tachyonic warm inflation model \cite{1-m}
\begin{eqnarray}\label{}
\nonumber
\Im(\phi)=-\int[\frac{(\frac{\Gamma}{V})'}{3H+\frac{\Gamma}{V}}+\frac{9}{8G}\frac{2H+\frac{\Gamma}{V}}{(3H+\frac{\Gamma}{V})^2}~~~~~~~~~~~~~~~~~~~~~~\\
\nonumber
\times[\Gamma+4HV-\frac{\Gamma'(\ln V)'}{36 H(3H+\frac{\Gamma}{V})}]\frac{(\ln V)'}{V}]d\phi.
\end{eqnarray}
The bulk viscous pressure effect leads to this difference.
Therefore, the perturbation parameters $P_R$, $R$, $n_s$ and $\alpha_s$  which may be found by WMAP and Planck observational data, for our model with viscous pressure, are modified due to the effect of this additional pressure.

\section{Exponential potential }
In this section we consider our model with the tachyonic
effective potential
\begin{equation}\label{46}
V(\phi)=V_0\exp(-\alpha\phi),
\end{equation}
where parameter $\alpha>0$ (with unit $m_p$) is related to mass
of the tachyon field \cite{8}. The exponential form of potential have
characteristics of tachyon field ($\frac{dV}{d\phi}<0,$ and
$V(\phi\rightarrow 0)\rightarrow V_{max}$ ). We develop our model
in high dissipative regime, i.e. $r\gg 1,$ for two cases: 1- $\Gamma$ and $\zeta$ are constant parameters, 2- $\Gamma$ as a function of tachyon  field $\phi$ and $\zeta$ as a function of energy density  $\rho$ of imperfect fluid.
\subsection{$\Gamma=\Gamma_0$, $\zeta=\zeta_0$ case}
From Eq.(\ref{39}), the slow-roll parameter $\tilde{\epsilon}$ in the present case has the form
\begin{equation}\label{47}
\tilde{\epsilon}=\frac{\sqrt{3}}{2}\frac{\alpha^2\sqrt{V_0}}{\Gamma_0}\exp(-\alpha\frac{\phi}{2}).
\end{equation}
%Also the other slow-roll parameter $\tilde{\eta}$ is obtained from Eq.(\ref{41})
%\begin{equation}\label{48}
%\tilde{\eta}=\frac{M_4^2}{8\pi}\frac{\alpha^2}{rV_0^2 e^{-2\alpha\phi}}
%\end{equation}
Dissipation parameter $r=\frac{\Gamma}{3HV}$ in this case is given by
\begin{equation}\label{49}
r=\frac{\Gamma_0}{\sqrt{3}V_0^{\frac{3}{2}}}\exp(\frac{3}{2}\alpha\phi)\gg 1.
\end{equation}
We find the evolution of tachyon field with the help of Eq.(\ref{7})
\begin{equation}\label{50}
\phi(t)=\frac{1}{\alpha}\ln[\frac{\alpha^2 V_0}{\Gamma_0}t+e^{\alpha\phi_i}],
\end{equation}
where $\phi_i=\phi(t=0)$.
Hubble parameter for our model has the form
\begin{equation}\label{51}
H=\sqrt{\frac{V_0}{3}}\exp(-\frac{\alpha\phi}{2}).
\end{equation}
At the end of inflation ($\tilde{\epsilon}\simeq 1$) the tachyon field becomes
\begin{equation}\label{52}
\phi_f=\frac{2}{\alpha}\ln[\frac{\sqrt{3V_0}\alpha^2}{2\Gamma_0}],
\end{equation}
so, by using the above equation and Eq.(\ref{50}) we may find time at which inflation ends
\begin{equation}\label{53}
t_f=\frac{3}{4}\frac{\alpha^2}{\Gamma_0}-\frac{\Gamma_0}{\alpha^2 V_0}e^{\alpha\phi_i}.
\end{equation}
Using Eqs.(\ref{12}) and (\ref{47}), the energy density of the radiation-matter fluid in high dissipative limit becomes
\begin{equation}\label{54}
\rho=\sqrt{\frac{V_0}{3\gamma^2}}\exp(-\frac{\alpha\phi}{2})[\frac{\alpha^2}{\Gamma_0}V_0\exp(-\alpha\phi)+3\zeta_0],
\end{equation}
and, in terms of tachyon field energy density $\rho_{\phi}$ becomes
\begin{equation}\label{55}
\rho=\frac{\rho_{\phi}^{\frac{1}{2}}}{\sqrt{3}\gamma}(\frac{\alpha^2}{\Gamma_0}\rho_{\phi}+3\zeta_0).
\end{equation}
For this example, the  entropy density in terms of energy density of inflaton may be obtained from above equation
\begin{equation}\label{}
Ts=\frac{\rho_{\phi}^{\frac{1}{2}}}{\sqrt{3}\gamma}(\frac{\alpha^2}{\Gamma_0}\rho_{\phi}+3\zeta_0).
\end{equation}
In FIG.1,  we plot the entropy density in terms of inflaton energy density. It may be seen that the entropy density increases by the bulk viscous effect \cite{mm-1}.
\begin{figure}[h]%=========================================================================== F1
\centering
  \includegraphics[width=10cm]{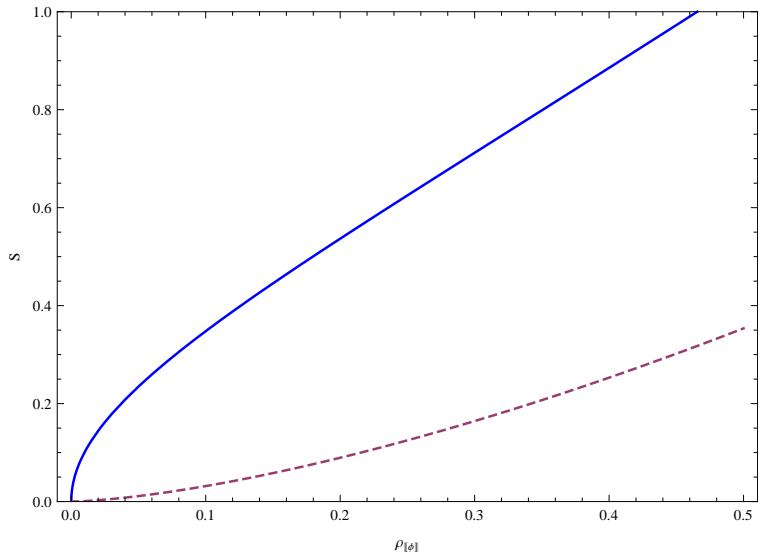}% Here is how to import EPS art
  \caption{We plot the entropy density $s$ in terms of energy density of tachyon field $\rho_{\phi}$ where, $\Pi=0$ by dashed curve and $\Pi=-3\zeta_0 H$ by blue curve ($T=5.47\times 10^{-5}$, $\Gamma=\Gamma_0=7.9\times 10^{3}, \zeta_0=4.21\times10^{-5}, \alpha=1$,$\gamma=\frac{4}{3}$). }
 \label{fig:F1}
\end{figure}
From Eq.(\ref{15}), the number of e-fold at the end of inflation, by using the potential (\ref{46}), for our inflation model is given by
%\begin{equation}\label{56}
%N_{total}=\sqrt{\frac{4\pi\lambda}{3M_4^2}}\frac{\Gamma_0}{\alpha}(\phi_f-\phi_i)
%\end{equation}
%or equivalently
\begin{equation}\label{57}
N_{total}=\frac{2\Gamma_0}{\alpha^2\sqrt{3V_0}}[\exp(\frac{\alpha\phi_f}{2})-\exp(\frac{\alpha\phi_i}{2})].
\end{equation}
where $\phi_f>\phi_i$. Using Eqs.(\ref{37}) and (\ref{45}), we could find the scalar spectrum and scalar-tensor ratio
\begin{equation}\label{58}
\delta_H^2=\frac{128\sqrt{\Gamma_0}}{225\sqrt[4]{3}\alpha^2}[\frac{V^2(\phi_0)}{(\sqrt{V(\phi_0)}+A)^{\frac{9}{2}}}]\frac{T_r}{\sqrt[4]{V(\phi_0)}},
\end{equation}
where $A=\frac{3\sqrt{3}\zeta_0}{8}(-3+\frac{5}{\gamma}),$
and
\begin{equation}\label{59}
R=\frac{9\sqrt[4]{3}}{5\sqrt{\Gamma_0}}\frac{(\sqrt{V(\phi_0)}+A)^{\frac{9}{2}}}{V^2(\phi_0)}\frac{V(\phi_0)^{\frac{5}{4}}}{T_r}\coth[\frac{k}{2T}],
\end{equation}
respectively, where the subscript zero $0$ denotes the time, when the perturbation was leaving
the horizon.
In the above equation we have used the Eq.(\ref{38}) where
\begin{equation}\label{}
\tilde{\Im}(\phi)=\ln(\frac{[\sqrt{V(\phi_0)}+A]^{\frac{9}{4}}}{V(\phi_0)}).
\end{equation}
These parameters may by restricted by WMAP9 and Planck data \cite{6,planck}. Based on these data, an upper bound for $V(\phi_0)$ may be found
\begin{equation}\label{}
\nonumber
V(\phi_0)<2.28\times 10^{-4}.
\end{equation}
In the above equation we have used these data: $R<0.11$, $P_R=2.28\times 10^{-9}$ \cite{6,planck}.
From Eqs.(\ref{va}) and (\ref{49}), non-Gaussianity for our model is presented as
\begin{equation}\label{}
f_{NL}=\frac{5\sqrt{3}}{9}\frac{\alpha^2 V_0^{\frac{3}{2}}}{\Gamma_0}\frac{\ln(\frac{\Gamma_0}{\sqrt{3}V_0^{\frac{3}{2}}})+\frac{3}{2}\alpha\phi(t_F)}{\exp(\frac{3}{2}\alpha\phi(t_F))},
\end{equation}
where the freeze-out time $t_F$ is the time when the last three wave-vectors $k$ thermalize \cite{1-m}.
\subsection{$\Gamma=\Gamma(\phi)$, $\zeta=\zeta(\rho)$ case}
Now we assume $\zeta=\zeta(\rho)=\zeta_1\rho,$ and $\Gamma=\Gamma(\phi)=\alpha_1 V(\phi)=\alpha_1 V_0\exp(-\alpha\phi)$, where $\alpha_1$ and $\zeta_1$ are positive constants. By using exponential potential (\ref{46}), Hubble parameter, $r$ parameter and slow-roll parameter $\tilde{\epsilon}$  have these forms
\begin{eqnarray}\label{60}
H(\phi)=\sqrt{\frac{V_0}{3}}\exp(-\frac{\alpha\phi}{2}),~~~~~~~~~~~~r=\frac{\alpha_1}{\sqrt{3V_0}}\exp(\alpha\frac{\phi}{2}),~~~~~~\\
\nonumber
\tilde{\epsilon}=\sqrt{\frac{3}{V_0}}\frac{\alpha^2}{2\alpha_1}\exp(\alpha\frac{\phi}{2}),~~~~~~~~~~~~~~~~~~~~~~~~~~~~~~~~~~~~~~~~~~
\end{eqnarray}
respectively. Using Eq.(\ref{7}), we find the scalar field $\phi$ in term of cosmic time
\begin{equation}\label{61}
\phi(t)=-\frac{\alpha}{\alpha_1}t+\phi_i.
\end{equation}
The energy density of imperfect fluid $\rho$ in terms of the inflaton energy density $\rho_{\phi},$ is given by the expression
\begin{equation}\label{62}
\rho=\frac{\alpha^2}{\alpha_1}\frac{\rho_{\phi}^{\frac{1}{2}}}{\sqrt{3}}(\gamma-\sqrt{3}\xi_1\rho_{\phi}^{\frac{1}{2}})^{-1}.
\end{equation}
We can find the entropy density $s$ in terms of energy density $\rho_{\phi}$
\begin{equation}\label{}
Ts=\frac{\alpha^2}{\alpha_1}\frac{\rho_{\phi}^{\frac{1}{2}}}{\sqrt{3}}(\gamma-\sqrt{3}\xi_1\rho_{\phi}^{\frac{1}{2}})^{-1}.
\end{equation}
The entropy density and matter-radiation energy density of our model in this case increase by the bulk viscosity effect (see FIG.2).
\begin{figure}[h]%=========================================================================== F1
\centering
  \includegraphics[width=10cm]{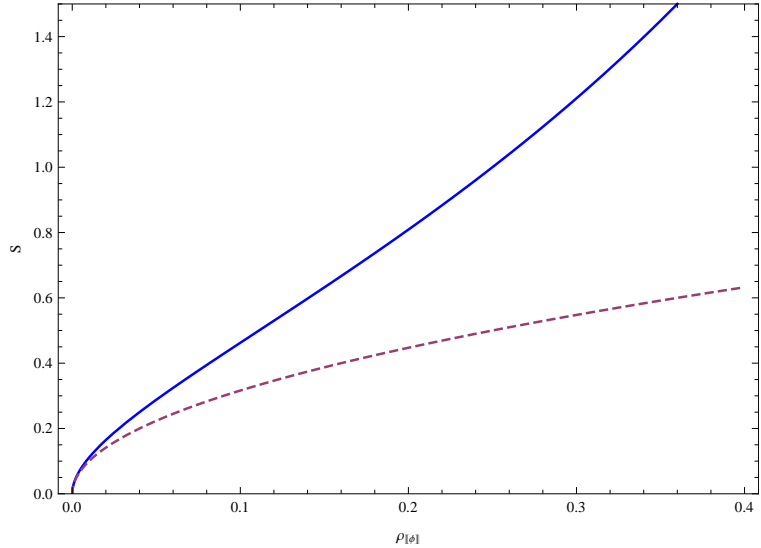}% Here is how to import EPS art
  \caption{We plot the entropy density $s$ in terms of scalar field energy density $\rho_{\phi}$ where, $\Pi=0$ by dashed line and $\Pi=-3\zeta_1\rho H$ by blue curve ($T=5.47\times 10^{-5},  \gamma=\frac{4}{3}, \alpha_1=1.38\times10^4,\zeta_1=0.76,\alpha=1$).}
 \label{fig:F2}
\end{figure}
From Eq.(\ref{61}) the scalar field and effective potential at the end of inflation where $\tilde{\epsilon}\simeq 1$, becomes
\begin{equation}\label{65}
\phi_f=\frac{1}{\alpha}\ln[\frac{V_0}{3}(\frac{2\alpha_1}{\alpha^2})^2],~~~~~~~V_f=\frac{3}{4}\frac{\alpha^4}{\alpha_1^2}.
\end{equation}
By using the above equation and Eq.(\ref{61}) we may find time at which inflation ends
\begin{equation}\label{}
t_f=\frac{3}{4}\frac{\alpha^2}{\Gamma_0}-\frac{\Gamma_0}{\alpha^2 V_0}e^{\alpha\phi_i}.
\end{equation}
Number of e-folds in this case is related to $V_i$ and $V_f$ by using Eq.(\ref{15})
\begin{equation}\label{63}
V_i=(2N-1)^2V_f.
\end{equation}
At the begining of the inflation $r$ parameter is given by
\begin{equation}\label{64}
r=r_i=\frac{2}{3}\frac{\alpha_1^2}{(2N-1)\alpha^2}.
\end{equation}
High dissipative condition($r\gg 1$), leads to $\alpha_1\gg\alpha(N-1)^{\frac{1}{2}}$ which is agree with the warm-tachyon inflation model without viscous pressure \cite{1-m}.
From Eqs.(\ref{va}) and (\ref{60}), the non-Gaussianity for our model in this case ($\Gamma=\Gamma(\phi)$, $\xi=\xi(\rho)$) is given by
\begin{equation}\label{}
f_{NL}=\frac{5\sqrt{3}}{9}\frac{\alpha^2\sqrt{V_0}}{\alpha_1}\frac{\ln(\frac{\alpha}{\sqrt{3V_0}})+\frac{\alpha\phi}{2}}{\exp(\frac{\alpha\phi}{2})}.
\end{equation}
By using  Eqs.(\ref{37}) and (\ref{45}) scalar power spectrum and tensor-scalar ratio result to be
\begin{eqnarray}\label{67}
\delta_H^2=\frac{128\sqrt{\alpha_1}}{225\sqrt[4]{3}\alpha^2}[\sqrt{V(\phi_0)}+B]^{-\frac{9}{2}(1+\frac{\zeta_1\alpha^2}{\gamma\alpha_1})}\\
\nonumber
\times\exp(\frac{9}{8}[\gamma-1]\frac{\alpha^2}{\sqrt{3}\gamma\alpha_1}V(\phi_0)^{\frac{1}{2}})\sqrt[4]{V(\phi_0)}T_r,
\end{eqnarray}
and
\begin{eqnarray}\label{68}
R=\frac{9\sqrt[4]{3}\alpha^2}{5\sqrt{\alpha_1}}[\sqrt{V(\phi_0)}+B]^{\frac{9}{2}(1+\frac{\zeta_1\alpha^2}{\gamma\alpha_1})}\\
\nonumber
\times\exp(-\frac{9}{8}[\gamma-1]\frac{\alpha^2}{\sqrt{3}\gamma\alpha_1}V(\phi_0)^{\frac{1}{2}})\frac{V(\phi_0)^{\frac{3}{4}}}{T_r},
\end{eqnarray}
respectively, where $B=\frac{3\sqrt{3}\zeta_1^2 V_0\alpha^2}{8\gamma\alpha_1}$.
In the above equations we have used the Eq.(\ref{38}) where
\begin{equation}\label{69}
\tilde{\Im}(\phi_0)=\frac{9}{16}(1-\gamma)(\frac{\alpha^2}{\sqrt{3}\gamma\alpha_1})
V^{-\frac{1}{2}}(\phi_0)+\frac{9}{4}[1+\frac{\zeta_1\alpha^2}{\gamma\alpha_1}]\ln(V^{\frac{1}{2}}(\phi_0)+B).
\end{equation}
These parameters may be restricted, using WMAP9, Planck and BICEP2 data \cite{6,planck,BICEP2}.
Using WMAP9 (BICEP2) data, $P_R(k_0)=\frac{25}{4}\delta_H^2\simeq 2.28\times 10^{-9}$, $R(k_0)\simeq 0.11,$($R(k_0)\simeq 0.22$) and the characteristic of warm inflation, $T>H$ \cite{3}, we may restrict the values of temperature
$T_r>5.47\times 10^{-5}$ ($T_r>7.73\times 10^{-5}$), using Eqs.(\ref{37}), (\ref{45}), or the corresponding equations (\ref{58}), (\ref{59}), (\ref{67}), (\ref{68}),
in our coupled examples, (see FIG.3). We have chosen $k_0=0.002 Mpc^{-1}$ and $T\simeq T_r$. Using BICEP2 data, we have found the new minimum of $T_r$ (See for example \cite{end}).
\begin{figure}[h]%=========================================================================== F1
\centering
  \includegraphics[width=10cm]{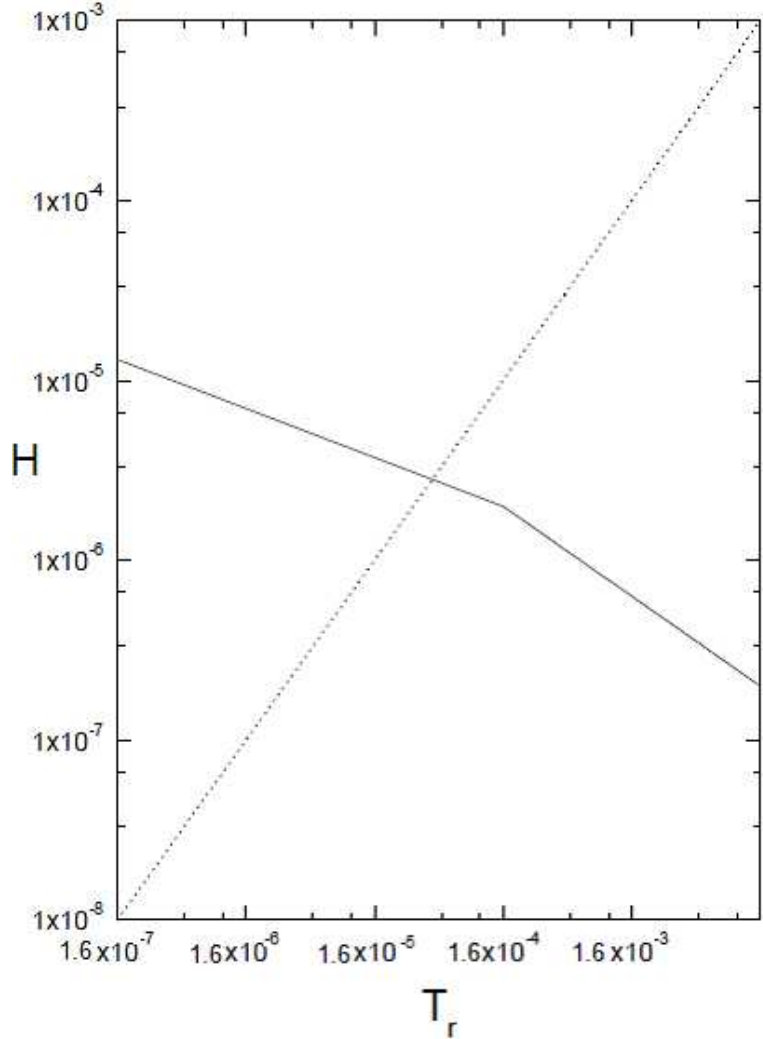}% Here is how to import EPS art
  \caption{In this graph we plot the Hubble parameter $H$ in term of the temperature $T_r$. We can find the minimum amount of temperature $T_r=5.47\times 10^{-5}$ in order to have the necessary condition for warm inflation model ($T_r>H$). }
 \label{fig:F3}
\end{figure}

%\begin{figure}
%\includegraphics{Vahid07.eps
%}
%\caption{We plot the parameter $m^2$ in term of dissipation parameter $\Gamma_0$ where $T=T_r=2.24\times 10^{16} GeV, K_{*}=0.002 Mpc^{-1} $ and $\kappa=1$  %}\label{fk3}
%\end{figure}
\section{Conclusion}
Warm-tachyon  inflation model with viscous pressure, using overlasting form of potential
$V(\phi)=V_0\exp(-\alpha\phi),$ which agrees with the tachyon
potential properties, has been studied. The main problem of the
inflation theory is how to attach the universe to the end of the
inflation period. One of the solutions of this problem is the
study of inflationary epoch in the context of warm inflation scenario \cite{3}. In
this model radiation is produced during inflation period where its
energy density is kept nearly constant. This is phenomenologically
fulfilled by introducing the dissipation term $\Gamma$.
Warm inflation model with viscous pressure is an extension of warm inflation model
where instead of radiation field we have radiation-matter fluid.
The study of warm inflation model with viscous pressure  as a mechanism that gives an end for
tachyon inflation are motivated us to consider the warm tachyon
inflation model with viscous pressure. In the slow-roll approximation the general relation between energy density of radiation-matter fluid and energy density of tachyon field is found. In longitudinal gauge and slow-roll limit the explicit expressions
for the tensor-scalar ratio $R$ scalar spectrum $P_R$ index,
$n_s$ and its running $\alpha_s$ have been obtained. We have
developed our specific model by exponential potential for two cases: 1- Constant dissipation coefficient $\Gamma_0$ and constant bulk viscous pressure coefficient $\zeta_0$. 2- $\Gamma$ as a function of tachyon field $\phi$ and $\zeta$ as a function of imperfect fluid energy density $\rho$ . In these two
cases we have found perturbation parameters and constrained these
parameters by WMAP9 and Planck data.

\end{document}